\documentclass[aps,prl,twocolumn,showpacs]{revtex4}
\usepackage{graphicx}
\usepackage{amssymb}

\begin{document}

\title{High-Curie-temperature ferrimagnetism and ferroelectricity in Bi$_2$FeMoO$_6$}

\author{Peng Chen}
\author{Bang-Gui Liu}
\email[Corresponding author:~]{bgliu@iphy.ac.cn}
\affiliation{Beijing National Laboratory for Condensed Matter
Physics, Institute of Physics, Chinese Academy of Sciences,
Beijing 100190, China}

\date{\today}

\begin{abstract}
BiFeO$_3$ is the most famous multiferroic material, but its G-type
antiferromagnetism is highly desirable to be replaced by strong
macroscopic magnetism beyond room temperature. Here we obtain double
perovskite Bi$_2$FeMoO$_6$ with R3 (\#146) space group by
substituting Mo for 50\% Fe in BiFeO$_3$. Our first-principles
calculated results show that it is a semiconductor with gap reaching
to 0.725 eV, its net magnetic moment is 2$\mu_B$ per formula unit,
and its ferroelectric polarization is 85$\mu$C/cm$^2$. This
ferroelctricity is comparable with that of BiFeO$_3$, but here the
magnetism is a strong ferrimagnetism with Curie temperature of 650
K. Our first-principles phonon spectra establishes that this R3
phase is stable. Electric polarization and magnetic easy axis are
shown to be in pseudo-cubic [111] axis. Our further analysis shows
that the multiferroic mechanism is similar to that in BiFeO$_3$.
Therefore, this Bi$_2$FeMoO$_6$ can be used to achieve strong
macroscopic magnetism and ferroelectricity well above room
temperature, being useful for designing new multifunctional
materials and devices.
\end{abstract}


\maketitle


Multiferroic materials have great potentials to achieve new
multifunctional devices for spintronics and data
storage\cite{r0,r1,r2,r3,r4}. Recent years witnesses that perovskite
BiFeO$_3$ becomes the most famous multiferroic
material\cite{r3,org1,org2,f3,f4,f5,f6,app1,app2,rnew,t1,t2,t3,ft1,ft2}.
It can show strong ferroelectricity with Curie temperature 1140K and
antiferromagnetism with magnetic Neel temperature
640K\cite{r3,f3,rnew}. In order to realize high-performance
applications, one hopes to achieve strong macroscopic magnetism
(ferromagnetism or ferrimagnetism) above room temperature instead of
antiferromagnetism in the BiFeO$_3$, keeping its strong
ferroelectricity\cite{r3,app1,app2,rnew}. For this purpose, it is a
big step forward to predict and then synthesize multiferroic
Bi$_2$FeCrO$_6$ whose double perovskite structure can be obtained
from perovskite BiFeO$_3$ by substituting Cr for 50\% Fe in the B
site and forming alternate Fe and Cr ordering along the [111]
axis\cite{fct1,fct2,fce1,fce2,fce3,fce4,fce5}. However, it is
difficult to keep the Fe and Cr ordering at a very high level
because of the small difference between Fe and Cr\cite{fce4,fce5}.
Therefore, it is highly desirable to realize much higher B-site
ordering and achieve much better materials with strong macroscopic
magnetism and ferroelectricity.

Here, we present a promising multiferroic Bi$_2$FeMoO$_6$ by
substituting Mo for 50\% Fe in perovskite BiFeO$_3$ and keeping
alternate order of Fe and Mo along the [111] axis. It is a double
perovskite phase with R3 (\#146) space group. Considering that
excellent Fe-Mo ordering has been acieved in double perovskite
Sr$_2$FeMoO$_6$, a well-known half-metallic
ferrimagnet\cite{add1,add2,add3}, we believe that this substitution
can lead to good ferrimagnetism because the moments at Fe and Mo
sites cannot completely compensate any more. Our systematic
first-principles calculations and structural analyses reveal that
double perovskite Bi$_2$FeMoO$_6$ has a semicinductor gap of 0.725
eV and can show both strong ferroelectricity and strong
ferrimagnetism with high Curie temperature. More detailed results
will be presented in the following.

\vspace{0.5cm}
\noindent \textbf{\large Results}\\
\textbf{Structure optimization.} Following Glazer¡¯s famous work on
perovskite structures\cite{g1,g2}, a systematic group
analysis\cite{g3,g4,g5} reveals possible crystal structures for a
specific double perovskite material with formula
$A_2$$B$$B^\prime$O$_6$. These group analyses can be used to
simplify our structural optimizations. Here, we consider only those
tilted $B$O6 and $B^\prime$O6 octahedra allown by group
theory\cite{g5}. For optimizing Bi$_2$FeMoO$_6$, we start with a
cubic perovskite structure with space group Fm$\bar{3}$m (\#225).
This corresponds to a$^0$a$^0$a$^0$ in Glazor's notation. The
optimization leads to rhombohedral lattice constant $a_c$=5.558\AA{}
and angle $\alpha_c$=60$^{\circ}$, primitive cell volume
$V_c$=121.41\AA{}$^3$, and total energy $E_c$=$-67.3366$eV. In this
structure, the O octahedra do not rotate around the [111] axis.
Allowing the O octahedra to able to rotate around the [111] axis, we
can obtain an optimized structure with R$\bar{3}$ (\#148)  space
group. It has rhombohedral lattice constant $a_1$=5.644\AA{} and
angle $\alpha_1$=61.63$^{\circ}$, primitive cell volume
$V_1$=131.76\AA{}$^3$, and total energy $E_1$=$-69.5888$eV. Clearly,
this symmetry breaking makes the total energy 2.2522 eV lower than
the cubic phase. This structure still keeps inverse symmetry and
therefore has no ferroelectricity, and in contrast with the cubic
phase, the O octahedra of Fe and Mo do rotate by +15.7$^{\circ}$ and
-17.4$^{\circ}$ around the [111] axis. This is a$^-$a$^-$a$^-$ tilt
(antiphase $R_4^+$ rotation) in Glazor's notation. For searching for
the ground-state phase of Bi$_2$FeMoO$_6$, it is useful to compare
it with BiFeO$_3$\cite{org2,t2,t3,rnew} and
Bi$_2$FeCrO$_6$\cite{fct1,fct2,fce1,fce2,fce3} whose ground state
phases are R3c (\#161) and R3 (\#146) phases, respectively. Because
\%50 Fe is substituted by Mo, there is no R3c structure for
Bi$_2$FeMoO$_6$ and we should turn to the R3 structure for possible
candidate of the ground-state phase of Bi$_2$FeMoO$_6$. For this
purpose, we allow the Fe ion to have a $\Gamma_4^-$ displacement
along the [111] direction\cite{t3,n50,n51}. This inverse symmetry
breaking makes the total energy move further downward when we
carefully optimize the structure. Consequently, we obtain a
rhombohedral structure with space group R3 (\#146). This optimized
structure has rhombohedral lattice constant $a_2$=5.743\AA{} and
angle $\alpha_2$=59.76$^{\circ}$, primitive cell volume
$V_2$=133.24\AA{}$^3$, and total energy $E_2$=$-69.8186$eV. This
energy is 0.2298 and 2.4820 eV lower than the R$\bar{3}$ (\#148) and
Fm$\bar{3}$m (\#225), respectively. Considering its similarity with
BiFeO$_3$ and Bi$_2$FeCrO$_6$, we believe that this R3 structure is
the ground-state phase for Bi$_2$FeMoO$_6$. Our analysis results
show that the tilting of oxygen octahedra of Fe and Mo can be
described with rotation angles of +17.8$^{\circ}$ and
-17.0$^{\circ}$. It is clear that the three-fold rotational symmetry
around the [111] axis is kept, but the calculated bond angles of
Fe-O-Mo (150.0$^{\circ}$), O-Fe-O (163.3$^{\circ}$), and O-Mo-O
(168.6$^{\circ}$) deviate substantially from 180$^{\circ}$. These
reveal that the oxygen octahedra are substantially deformed. We
illustrate the crystal structure and local octahedral tilting and
deformation in Fig. 1 and summarize our calculated parameters in
Table I. Although we have let only the Fe ion move along the [111]
direction with respect to its oxygen environment in the initial
structure, the optimization makes all the four cations coherently
move along the [111] direction with respect to their oxygen
environments, and we shall show that these coherent displacements
are favorable to form ferroelectricity.

\begin{figure}[!htbp]
\includegraphics[width=7cm]{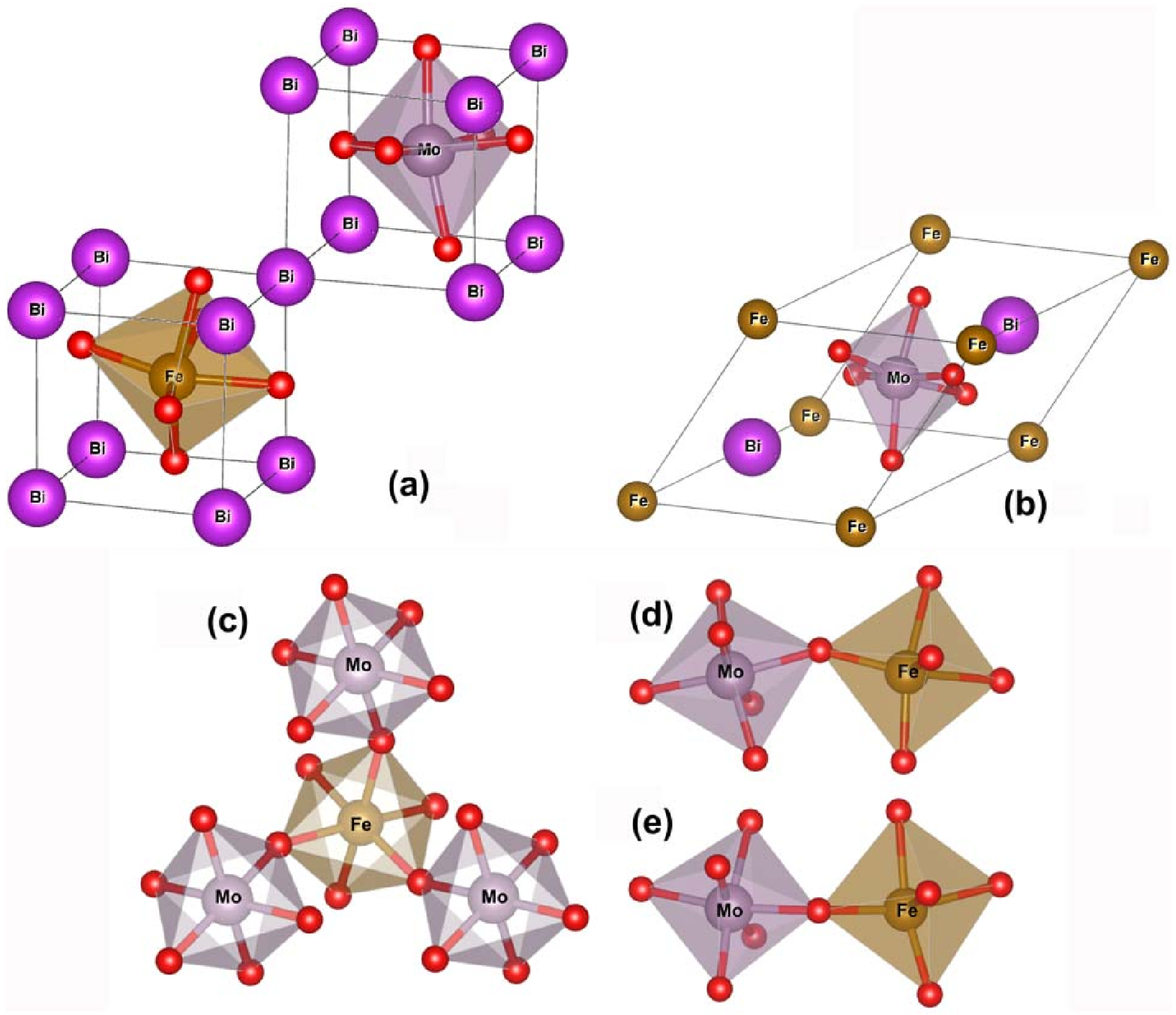}
\caption{(Color online.) Crystal structure (a,b) and local changes
of O octahedra of Fe and Mo (c-e) of double perovskite
Bi$_2$FeMoO$_6$ with the \#146 space group. (a) A schematic crystal
structure, showing the alternate occupation of Fe and Mo at the
B-sites of the simple perovskite structure along the [111]
direction. (b) The primitive cell of the double perovskite structure
with the O octahedron of Mo shown. (c) The relative rotation
projected on the (111) plane of an O octahedron of Fe with respect
to three of the nearest O octahedra of Mo. (d) The two neighboring O
octahedra of Fe and Mo projected on the plane of the Fe-O-Mo
triangle, reflecting the Fe-O-Mo bond angle. (e) the same two O
octahedra of Fe and Mo projected on the perpendicular plane
including the Fe and Mo ions, showing local deformation.}
\label{fig:1}
\end{figure}

\begin{table}[!htb]
\caption{Main parameters of the fully optimized double perovskite
Bi$_2$FeMoO$_6$ with the \#146 space group: rhombohedral lattice
constant ($a$) and angle ($\alpha$), relative coordinates of the
different ions, and rotational angles of the O octahedron of Fe
($\Delta \theta_{\rm Fe}$) and Mo ($\Delta \theta_{\rm Mo}$).}
\label{tab:1}
\begin{ruledtabular}
\begin{tabular}{lllllllll}
& $a$  &&   5.743 \AA  &&  $\alpha$ && 59.76$^\circ$ & \\ \hline
\hline
& Bi&&     0.2244  &&  0.2244 &&   0.2244 &\\
&   &&     0.7289  &&  0.7289 &&   0.7289 &\\ \hline & Fe&&
0.0000  &&  0.0000 &&   0.0000 &\\ \hline & Mo&&     0.5089  &&
0.5089 &&   0.5089 &\\ \hline
& O &&     0.8432  &&  0.6774 &&   0.2913 &\\
&   &&     0.2913  &&  0.8432 &&   0.6774 &\\
&   &&     0.6774  &&  0.2913 &&   0.8432 &\\
&   &&     0.1765  &&  0.3511 &&   0.7917 &\\
&   &&     0.7917  &&  0.1765 &&   0.3511 &\\
&   &&     0.3511  &&  0.7917 && 0.1765 &\\\hline \hline &
$\Delta\theta_{\rm Fe}$ &&  17.8$^\circ$  && $\Delta\theta_{\rm Mo}$
&& -17.0$^\circ$ & \\
\end{tabular}
\end{ruledtabular}
\end{table}

\textbf{Magnetic moment and electronic energy gap.} Presented in
Fig. 2 is our GGA calculated spin-dependent density of states (DOS)
and energy bands of the optimized double perovskite Bi$_2$FeMoO$_6$
(R3 phase). Our analysis indicates that in the spin-up channel, the
three filled Fe t2g bands merges into the oxygen p bands and the two
filled Fe eg bands are separated from them, and the empty bands,
from Mo 4d states, are above 0.65 eV. It can be seen that there is a
narrow semiconductor gap, 0.059 eV, which is formed between the
three filled Mo 4d bands and the three empty Fe 3d ones in the
spin-down channel. This narrow GGA gap remains open as long as the
optimization is done with high force convergence standard of 1
meV/\AA{} or smaller, or else it can become a pseudo-gap indicating
a metallic electron phase\cite{lsd}. Totally, we have five Fe d
electrons in the spin-up channel and three Mo d electrons in the
spin-down channel, and thus the magnetic moment per formula unit is
equivalent to 2$\mu_B$, which is consistent with the calculated
data. The small gap value is due to the well-known under-estimation
of GGA on semiconductor gaps. In order to make an accurate
calculation of the semiconductor gap, we use a modified
Becke-Johnson (mBJ) exchange functional to replace the GGA one. Our
mBJ calculated gap, 0.725 eV, should be a much better value for the
true gap because mBJ has been proved to give accurate semiconductor
gaps for most of semiconductors. We also use other
exchange-correlation schemes to calculate the semiconductor gap and
conclude that there is certainly a finite semiconductor gap for the
double perovskite Bi$_2$FeMoO$_6$.

\begin{figure}[!htbp]
\includegraphics[width=7.6cm]{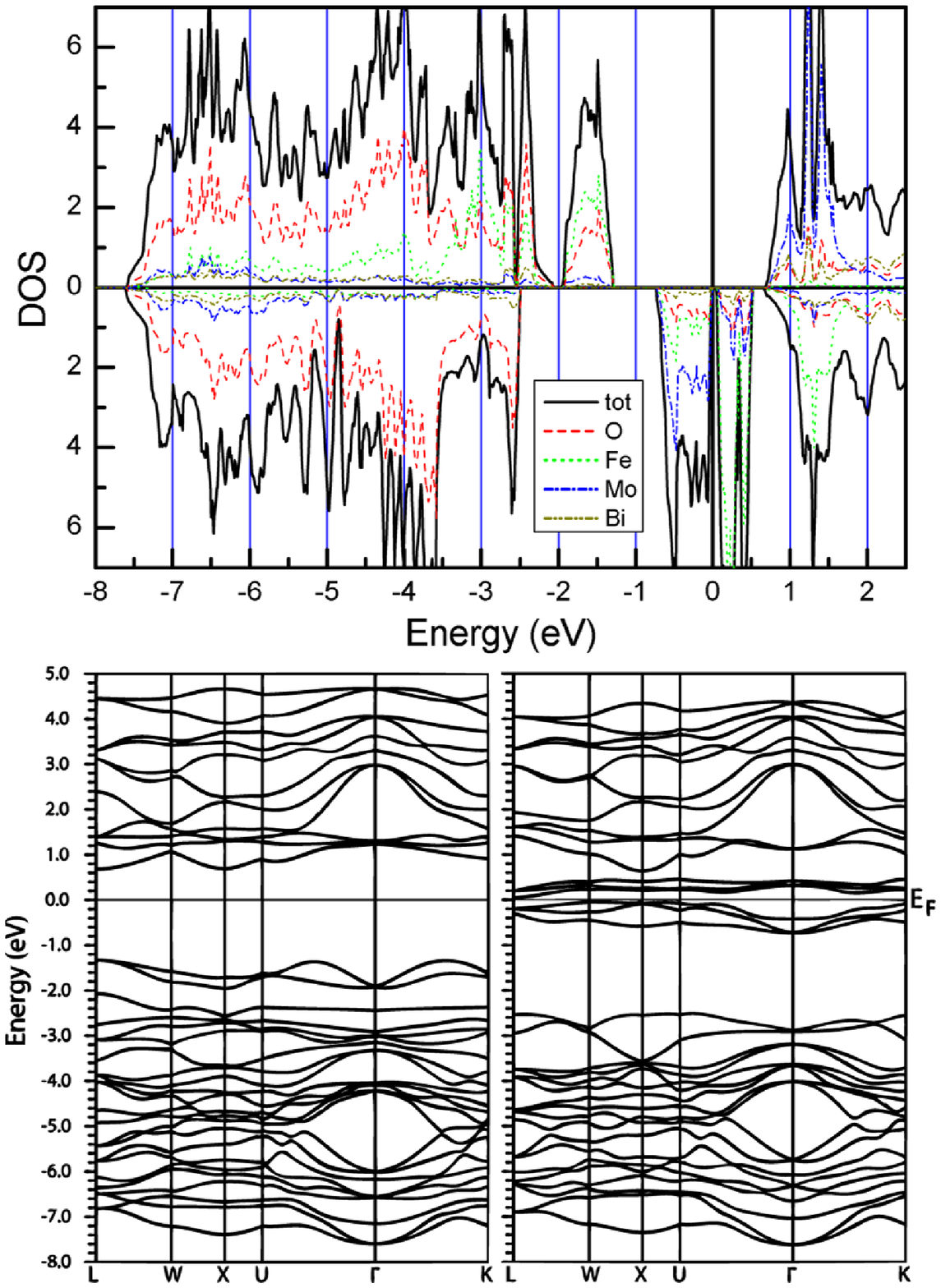}
\caption{Electronic structure of the double perovskite
Bi$_2$FeMoO$_6$ calculated with GGA. Upper: the spin-resolved
density of states (DOS, state/eV per primitive cell), in which the
upper part are spin-up DOS and the lower part spin-down DOS. Lower:
the spin-resolved energy band structures, in which the left part
shows spin-up bands and the lower part spin-down ones.}
\label{fig:3}
\end{figure}

\textbf{Ionic displacements towards ferroelectricity.} In Fig. 3 we
shows the relative distances between the cations and their nearest O
environment along the [111] direction for the three structures:
\#225, \#148, and \#146. The vertical line, $\Delta_1$-$\Delta_4$,
indicates the position of three O atoms forming a regular triangle
perpendicular to the [111] axis. The O-O bond lengthes and rotation
angles of the O octahedron are presented at the top of vertical
lines. The lengthes of the primitive cells along [111] direction are
given at the right-most sides of the [111] lines. Comparing the
\#148 structure with the \#225, Fe and Mo are still at the centers
of their O octahedra, respectively, and the local asymmetry at Bi1
and Bi2 ions in the \#148 structure becomes much larger than in the
latter. However, the displacements of Bi1 and Bi2 with respect to
their O environments are opposite to each other along the [111]
axis, which means that there is still no ferroelectricity, but an
Bi-induced antiferroelectricity, in the \#148 structure. In
contrast, the \#146 structure is very different in that all its
cations have in-phase displacements in the same [111] direction with
respect to their O environment. In order to get a clearer insight,
we further investigate the displacements of cations with respect to
their O environments along the [111] direction in terms of two
cation-related quantities, Plus and Minus. The Plus (Minus) value of
a cation is defined as the summation (subtraction) of the two
distances from the cation to its left O triangle and right triangle
along the [111] axis. The Plus value indicates the distance between
the two O triangles on the two sides of the cation, and the Minus
value describes the displacement of the cation with respect to its
two O triangles along the [111] direction. We present the Plus and
Minus values of the four cations in the \#146 phase of the
Bi$_2$FeMoO$_6$ in Table II. The Minus values of all the four
cations are negative and thus in-phase, and therefore indicate
ferroelectricity in the Bi$_2$FeMoO$_6$, being the same as those of
the Bi$_2$FeCrO$_6$ which has a strong ferroelectricity along the
[111] direction\cite{fct1,fct2,fce1,fce2,fce3,fce4,szw}.

\begin{figure*}[!htbp]
\includegraphics[width=14cm]{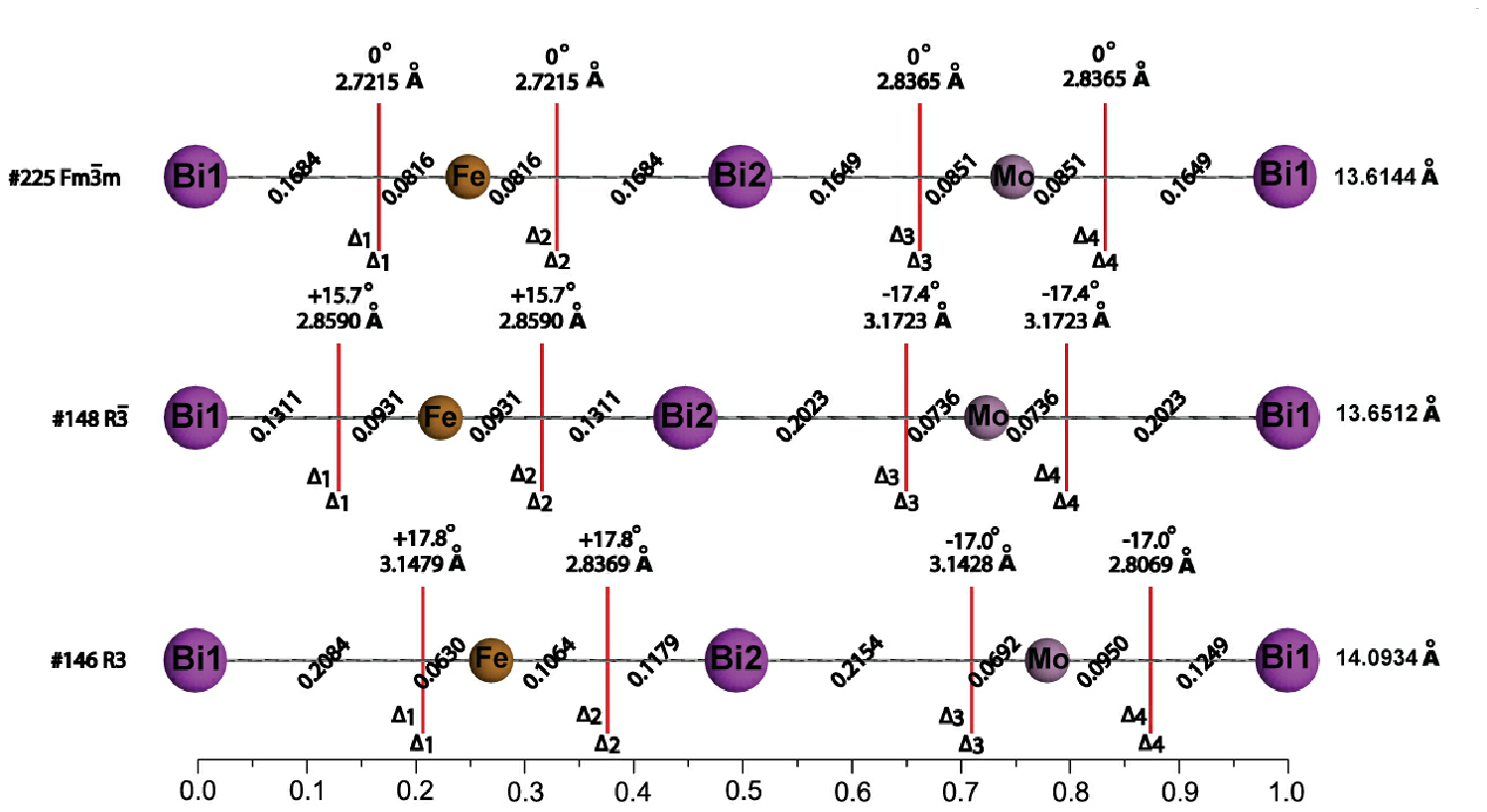}
\caption{(Color online.) The relative distances (tilted numbers)
between the positive ions (Bi1, Fe, Bi2, Mo) and the centers of the
nearest O triangles ($\Delta_1$, $\Delta_2$, $\Delta_3$, $\Delta_4$)
and the relative rotational angles of the O triangles around the
[111] axis, compared among the three crystal structures with the
space groups: \#225, \#148, and \#146. The values in \AA{} above the
O triangles are their edge lengthes (O-O distances), respectively.
The three values at the rightmost side are the lengthes of the
primitive cells along the [111] direction. Notice: there is a
rotational angle of $60^{\circ}$ between $\Delta_1$ and $\Delta_2$,
and  between $\Delta_3$ and $\Delta_4$.} \label{fig:4}
\end{figure*}

\begin{figure*}[!htbp]
\includegraphics[width=14cm]{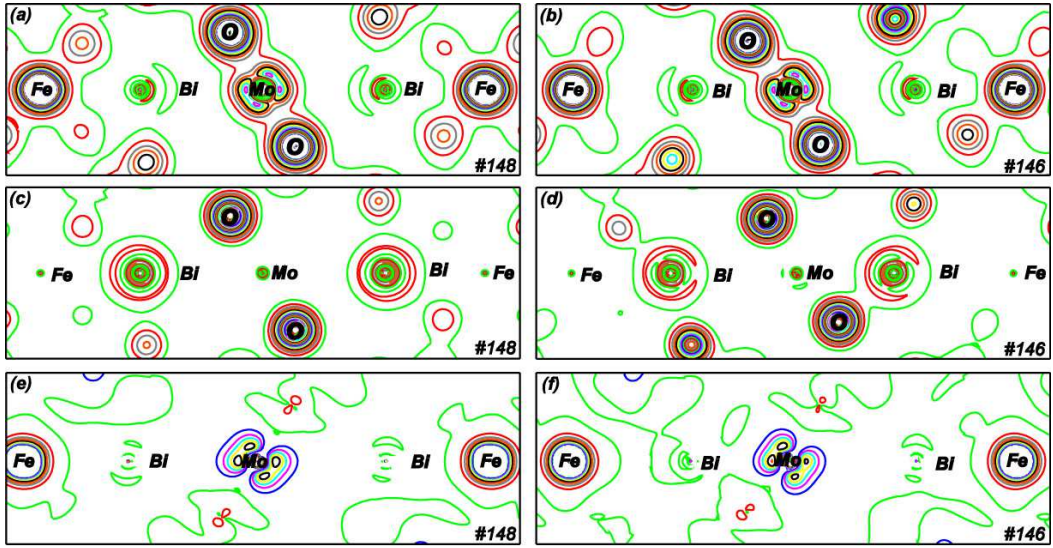}
\caption{(Color online.) The electron density distributions in the A
plane of the double perovskite Bi$_2$FeMoO$_6$ with the \#148 [left:
(a), (c), (e)] and \#146 [right: (b), (d), (f)] space groups in
comparison. (a) and (b) show the charge density distributions of the
valence electrons between -8 and 0 eV, ranging from 0 to
3$|e|$/\AA$^3$ with a contour increment of 0.15$|e|$/\AA$^3$; (c)
and (d) the charge density distribution of Bi 6s and O 2s electrons
between -21 and -9 eV, ranging from 0 to 2$|e|$/\AA$^3$ with a
contour increment of 0.1$|e|$/\AA$^3$; and (e) and (f) the spin
density distributions of the valence electrons between -8 and 0 eV,
ranging from 0 to 2$\mu_B$/\AA$^3$ with a contour increment of
0.1$\mu_B$/\AA$^3$.} \label{fig:5}
\end{figure*}

\textbf{Electron density distributions implying ferroelectricity.}
We present in Fig. 4 the electron density distributions in a typical
plane for the \#148 and \#146 structures of the Bi$_2$FeMoO$_6$. The
plane, namely A plane, is defined to include the Fe-Bi-Mo-Bi line
along the [111] axis and the two nearest O atoms on the two sides of
the Mo atom. The charge and spin density distributions can be
compared between the two structures. An asymmetry can be seen near
the two Bi atoms for the charge density distribution between -8 and
0 eV for the two structures, and it is, however, in phase for the
\#146 structure but out of phase for the other. Furthermore, we can
easily see an in-phase asymmetry in the Bi 6s charge density
distribution in the \#146 structure, but cannot in the other. As for
the spin density distribution, a difference can also be found
between the two Bi atoms for the \#146 structure. These electron
density properties are consistent with the in-phase displacements of
the cations in the [111] direction.

\textbf{DFT calculations of ferroelectric polarization.} Finally, we
quantitatively investigate the ferroelectricity of the
Bi$_2$FeMoO$_6$ through modern DFT-based calculations\cite{berry}.
Since the R$\bar{3}$ (\#148) as the reference structure has no
semiconductor gap under GGA, we make quantitative calculations of
the electric polarization by using GGA+U\cite{ldau1,ldau2} and
HSE06\cite{hse1,hse2} methods which have been implemented in VASP
package. For the GGA+U scheme, we use three sets of ($U$,$J$)
parameters for Fe, keeping the same ($U$,$J$)=(2.5eV,0.5eV) for Mo.
With these schemes and parameters, the semiconductor gaps range from
0.23 to 0.91 eV, but it should be kept in mind that accurate
semiconductor gap needs to be calculated with mBJ scheme. We
summarize the calculational parameters and electric polarization
results in Table II. These results means that the Bi$_2$FeMoO$_6$
(\#146) has a large spontaneous ferroelectric polarization
85$\mu$C/cm$^2$, which is almost independent of calculational
schemes and parameters. Therefore the Bi$_2$FeMoO$_6$ is an
excellent ferroelectric material.

\begin{figure}[!htbp]
\includegraphics[width=7.6cm]{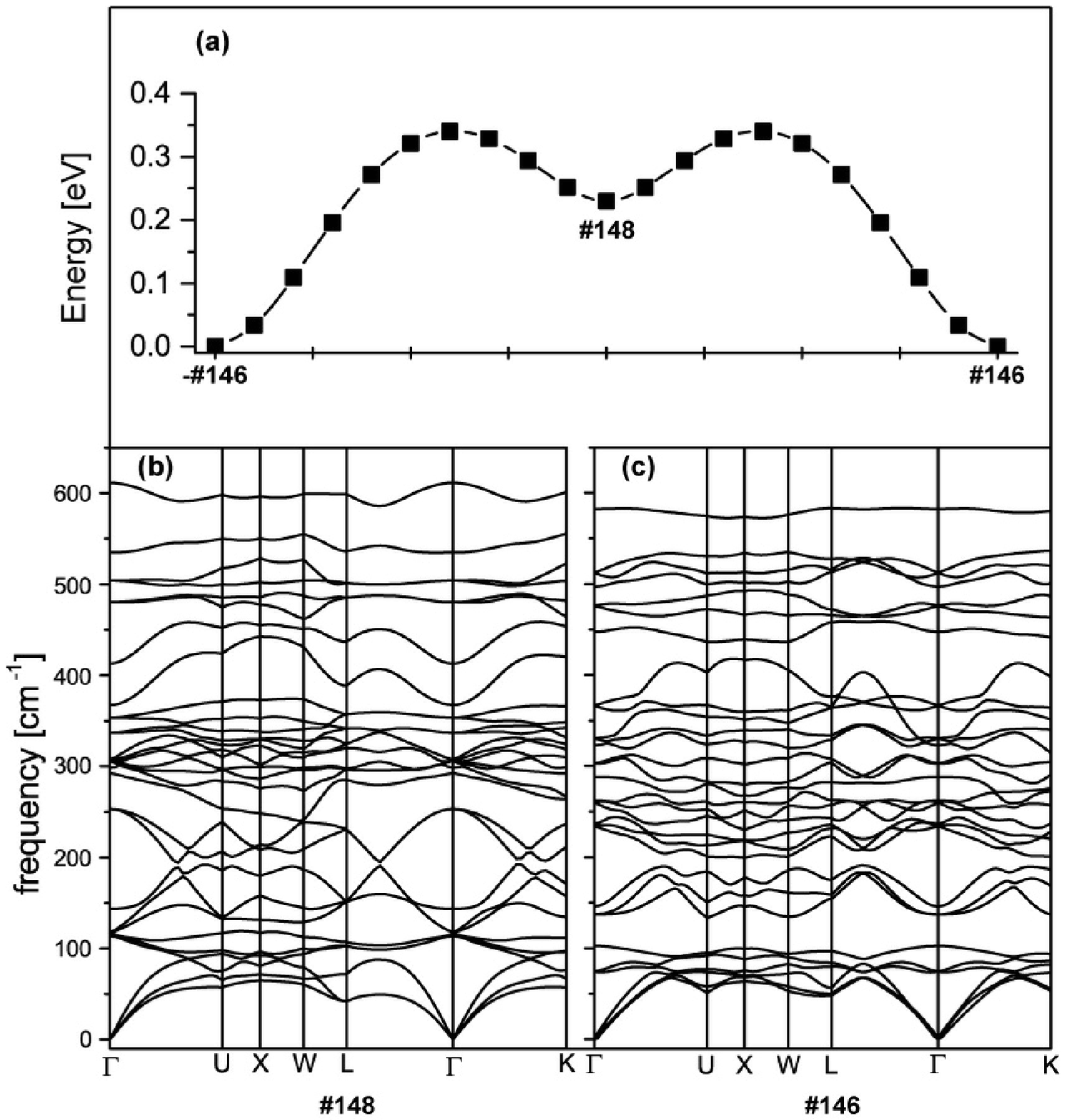}
\caption{(Color online.) Polarization reversal path (a) and
structural stability (b,c). (a) A reversal path of the ferroelectric
phase of \#146 through the intermediate phase of \#148. (b) Phonon
spectra of the intermediate phase (\#148). (c) Phonon spectra of the
ferroelectric phase (\#146).} \label{fig:2}
\end{figure}

\textbf{Ferroelectric switching and phonon spectra.} It should be
pointed out that the R3 (\#146) phase is degenerate in total energy.
For example, one of them, with the polarization $P$, can be changed
into another one with the polarization $-P$. Using a linear
interpolation method and performing full optimization on the basis
of the two with $\pm P$, we can obtain a reversal path connecting
them, with the R$\bar{3}$ (\#148) phase naturally appearing as an
intermediate structure. The calculated energy curve along the
path\cite{vasp1,vasp2} is shown in Fig. 5(a). Most importantly, we
have performed first-principles calculations on phonon spectra for
these two phases. The calculated phonon spectra are presented in
Fig. 5 (b) and (c). These results show that both of them are
dynamically stable, in addition to the static stability of the R3
phase in terms of formation heat calculations\cite{lsd}.

\begin{figure}[!htbp]
\includegraphics[width=6cm]{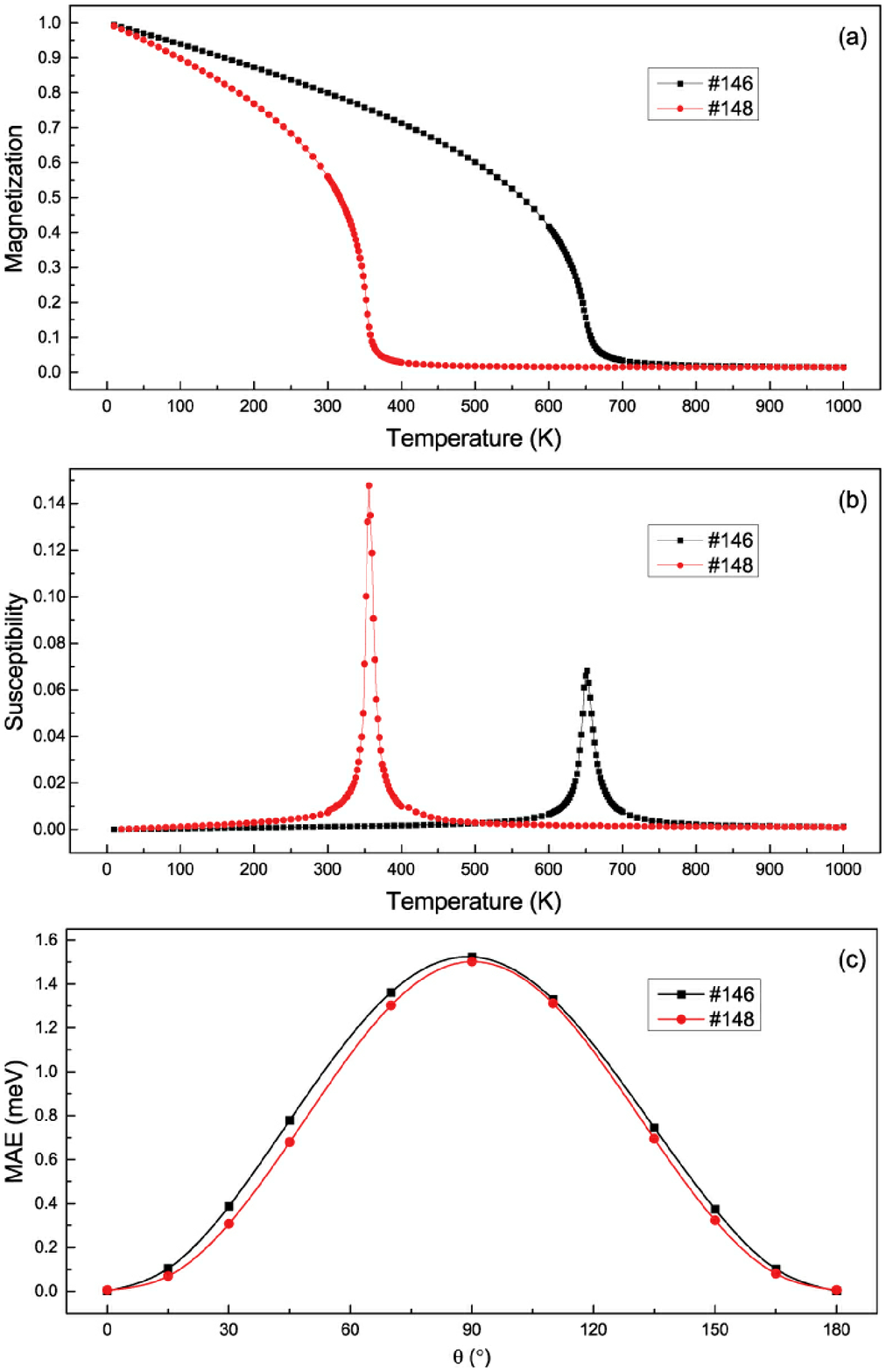}
\caption{(Color online.) The magnetism of the ferroelectric phase
and the intermediate phase of the double perovskite Bi$_2$FeMoO$_6$.
Monte Carlo simulated results for magnetization (a) and
susceptibility (b) of the two phases. (c) The magnetocrystalline
energies of the two phases, defined as a functions of the polar
angle (in degree) from the [111] direction.} \label{fig:6}
\end{figure}

\textbf{Robust multiferroicity.} All our above analyses of the ion
displacements and electronic density distributions indicate that the
double perovskite Bi$_2$FeMoO$_6$ in the \#146 structure should have
a strong ferroelectricity in the pseudocubic [111] direction. More
importantly, our DFT-based Berry phase calculation establishes that
it indeed has strong ferroelectricity $P$ along the direction. Its
electric polarization, reaching to 85$\mu$C/cm$^2$, is comparable
with that of the BiFeO$_3$, which indicates that it should have a
high ferroelectric Curie temperature. In addition, our analysis
indicate that there is another degenerate ground-state phase with
ferroelectricity $-P$, which implies that the polarization can be
reversed through the \#148 structure as an intermediate structure in
between. On the other hand, we also calculate spin exchange energies
and Curie temperatures for the multiferroic phase and the
intermediate one. The simulated curves for magnetization and
susceptibility are presented in Fig. 6 (a) and (b), indicating a
high magnetic Curie temperature of 650 K\cite{lsd}. Our
magnetocrystalline calculations including the spin-orbit coupling
are presented in Fig. 6(c). The curve for the \#146 phase can be
fitted with $E=E_0+E_{\rm MA}\sin^2(\frac{\pi}{180}\omega)$, with
$E_{\rm MA}=1.52\pm 0.005$ meV. This indicates that the easy axis is
the pseudocubic [111] and the magneto-crystalline anisotropic energy
is $1.52$ meV. It is interesting that the \#148 phase also has
ferrimagnetic order with a net magnetic moment of 2 $\mu_B$ below
360 K and magnetocryslline energy 1.50 meV with the easy axis long
the [111]. For convenience, we also summarize the key multiferroic
parameters in Table II. Therefore, it has been established that the
double perovskite Bi$_2$FeMoO$_6$ is a robust multiferroic material
with high Curie temperatures.

\begin{table}[!htb]
\caption{Calculated Plus and Minus values (\AA) of cations,
DFT-calculated ferroelectric polarizations, and summarized key
multiferroic parameters of the double perovskite Bi$_2$FeMoO$_6$.}
\label{tab:2}
\begin{ruledtabular}
\begin{tabular}{ccccc}
\multicolumn{5}{c}{Ionic displacement analysis}\\ \hline
 &    Bi1    &   Fe   &   Bi2   &   Mo \\ \hline
 Plus & 4.7337  & 2.3836 &  4.6978 & 2.3142 \\
Minus & -1.1771  & -0.6164 &  -1.3736 & -0.3646 \\ \hline \hline
\multicolumn{5}{c}{Berry phase calculations of polarization}\\
\hline
 scheme & \multicolumn{3}{c}{Fe($U$,$J$) / Mo($U$,$J$) (eV)} & $P$ ($\mu$C/cm$^2$) \\\hline
 GGA+U & \multicolumn{3}{c}{(2.0,0.8) / (2.5,0.5)} & 85.4  \\
 &  \multicolumn{3}{c}{(4.0,0.8) / (2.5,0.5)}  & 85.5  \\
 &  \multicolumn{3}{c}{(8.0,0.8) / (2.5,0.5)}  & 84.7  \\ \hline
 HSE06 & \multicolumn{3}{c}{-}  & 85.4  \\ \hline \hline
\multicolumn{5}{c}{Key multiferroic parameters}\\ \hline
 Ferrimagnetism & $M$ & 2 $\mu_B$ & $T_c$ & 650 K \\ \hline
 Ferroelectricity & $P$ & 85 $\mu$C/cm$^2$ & $T_c$ & high \\
\end{tabular}
\end{ruledtabular}
\end{table}

\vspace{0.5cm}
\noindent \textbf{\large Discussion}\\
We have shown robust ferrimagnetism and ferroelectricity in the
double perovskite Bi$_2$FeMoO$_6$ in the R3 (\#146) structure. The
large ferroelectric polarization in Table II means that this
Bi$_2$FeMoO$_6$ is comparable with the BiFeO$_3$ in
ferroelectricity, and in addition the strong ferrimagnetism is shown
to be able to persist at high temperature far beyond room
temperature. Because the Bi$_2$FeMoO$_6$ is stable in terms of both
static and dynamical calculations, it could be synthesized soon.
When realized experimentally, it should be a promising multiferroic
material, especially for spintronics applications. Because of its
robust multiferroic properties, the Bi$_2$FeMoO$_6$ can be used to
form some interesting composite structures such as multiferroic
superlattices\cite{zwy}.

As for the mechanism of multiferroicity, it is similar to that of
the BiFeO$_3$, except that here the Mo moment cannot be completely
compensated by the Fe moment, producing the finite net magnetic
moment 2$\mu_B$ per formula unit. It is interesting that the
ferrimagnetic Curie temperature is predicted to be also similar to
the Neel temperature of the BiFeO$_3$. The ferri-distortive
rotations of the oxygen octahedra of Fe and Mo is favorable to
stabilize the Bi-dominated antiferroelectricity in the \#148
structure of the Bi$_2$FeMoO$_6$. These ferri-distortive rotations
of the O octahedra of Fe and Mo are similar to antiferrodistortive
rotations of O octahedra of Fe in the BiFeO$_3$\cite{n50}. It is
also clear that these ferri-distortive rotations coexist with the
ferroelectricity in the \#146 phase. It should be pointed out that
in addition to the the multiferroic \#146 phase, there should exist
some metastable phases for Bi$_2$FeMoO$_6$, with the \#148 structure
being one of them. This is also similar to BiFeO$_3$ for which many
metastable structures have been found\cite{rnew,t3,ft1,ft2,n51}.

In summary, we have investigated double perovskite Bi$_2$FeMoO$_6$
through systematic DFT calculations and structural analyses. Our
careful GGA optimization reveals that the Bi$_2$FeMoO$_6$ in the
\#146 structure is a ferrimagnetic semiconductor. The net moment
reaches to 2$\mu_B$ per formula unit because the magnetic moments of
Fe and Mo cannot be canceled, and the semiconductor gap corrected
with mBJ potential reaches to 0.725 eV. The O octahedra of Fe and Mo
ions have rotation angles of +17.8$^{\circ}$ and -17.0$^{\circ}$
around the [111] axis, respectively. Our analyses of the electronic
structure and ionic displacements indicate strong ferroelectricity
in the Bi$_2$FeMoO$_6$. Our DFT-based electric polarization
calculations directly shows that the Bi$_2$FeMoO$_6$ has strong
ferroelectricity in the [111] direction. Our first-principles phonon
spectra show that this multiferroic phase is stable. Therefore, it
has been established that the Bi$_2$FeMoO$_6$ is a promising
multiferroic material with strong ferroelectricity and high
ferrimagnetic Curie temperature far beyond room temperature. With
the spin-orbit effect into account, our calculation show that the
magnetic easy axis is also in the [111] axis. The magnetocrystalline
anisotropy should be manipulated as has been done in Sr$_2$FeMoO$_6$
film materials\cite{add3}. This multiferroic material can be useful
for designing new multifunctional materials and devices in the
future.

\vspace{0.5cm}
\noindent \textbf{\large Methods}\\
We use projector augmented-wave (PAW)\cite{paw} plus
pseudo-potential methods within the density functional
theory\cite{dft,lda}, as implemented in the Vienna ab initio
simulation package (VASP) \cite{vasp1,vasp2}, to optimize the
crystal structures and then study the electronic structures,
ferroelectricity, and magnetic properties. The spin-polarized
generalized-gradient approximation (GGA)\cite{pbe}, to the
electronic exchange-correlation functional, is used to do structure
optimization. We use a plane wave basis set with a maximum kinetic
energy of 500 eV and a $12\times 12\times 12$ k-point mesh generated
with the Monkhorst-Pack scheme\cite{mp}. During optimization, all of
the structures were fully relaxed until the largest force between
the atoms become less than 1 meV/\AA{}.

Phonon spectra are calculated with first-principles perturbation
method with norm-conserving pseudo-potentials as implemented in
package Quantum-ESPRESSO \cite{phonon}. The magnetic anisotropy
energy (MAE) due to the spin-orbit coupling is determined by the
force theorem, and calculated in terms of the total energy with
respect to the [111] direction. We use a modified Becke-Johnson
exchange potential to accurately calculate semiconductor
gaps\cite{mbj}. We use modern Berry phase method\cite{berry} and
take both GGA+U \cite{ldau1,ldau2} and HSE06 \cite{hse1,hse2}
schemes to calculate electric polarization. For calculating magnetic
Curie temperatures, we first construct an effective Heisenberg spin
model through comparing total energies of the ferrimagnetic phase
and other magnetic structures concerned\cite{lsd}, and then do
standard Monte Carlo simulations in terms of the model.

\vspace{0.5cm}
\noindent \textbf{\large Acknowledgments}\\
This work is supported by Nature Science Foundation of China (Grant
No. 11174359), by Chinese Department of Science and Technology
(Grant No. 2012CB932302), and by the Strategic Priority Research
Program of the Chinese Academy of Sciences (Grant No. XDB07000000).
PC is grateful to Sai Gong for some technical helps.

\vspace{0.5cm}
\noindent \textbf{\large Author contributions}\\
BGL conceived and supervised the project, and PC carried out all the
calculations. PC and BGL both contributed to the analysis and
discussions of the results, and BGL and PC wrote the manuscript.

\vspace{0.5cm}
\noindent \textbf{\large Additional information}\\
Competing financial interests: The authors declare no competing
financial interests.

\end{document}